\documentclass[twoside]{LCWS11}
\usepackage[latin1]{inputenc}
\usepackage[dvips]{graphicx,epsfig,color}
\usepackage{wrapfig,rotating}
\usepackage{amssymb,amsmath,array}
\usepackage{cite}
\newcommand{\kakeru}{$\times$}
\newcommand{\SOL}{$c$}

\newcommand{\celd}{$^\circ$C}

\begin{document}
\title{
%%%%   Paper title goes here  %%%%%%%%%%%%%%
Performance of the CALICE Scintillator-Based ECAL Depending on the Temperature } %% 
%***********************************************************************
% AUTHORS INFORMATION AREA
%***********************************************************************
%\author{Katsushige Kotera$^1$ and Adil Khan$^2$ on behalf of the CALICE collaboration
\author{Katsushige Kotera on behalf of the CALICE collaboration
% Optional short acknowledgment: remove next line if non-needed
%\thanks{This is an optional funding source acknowledgment.}
% DO NOT MODIFY THE FOLLOWING '\vspace' ARGUMENT
\vspace{.3cm}\\
% Addresses and institutions (remove "1- " in case of a single institution)
Shinshu University - Faculty of Science \\
Asahi-3-1-1, Matsumoto, Nagano - Japan
%% Remove the next three lines in case of a single institution
\vspace{.1cm}\\
%2- Kyungpook National University - Center For High Energy Physics Department \\
%Daegu - S - Korea\\
}
%%***********************************************************************
% END OF AUTHORS INFORMATION AREA
%***********************************************************************

\maketitle

\begin{abstract}
The CALICE collaboration is developing a granular electromagnetic calorimeter using (5-10)\,\kakeru\,45\,\kakeru\,3\,mm$^3$ scintillator strips for a future  linear collider experiment.
Each scintillator strip is read out by using a Pixelated Photon Detector (PPD).
A prototype module has been constructed and tested at Fermilab in 2008 and 2009.  
Since the sensitivity of the PPD is affected by temperature fluctuations, a temperature correction method has been established which performed such that 
the deviation of the measured response from a linear behavior is improved from greater than 10\% to less than 1.5\%.
\end{abstract}

\section{Introduction}

%%%%%%%%%%%%%%%%%%%%%%%%%%%%%%%%%%
%%%%%%%%%%%%%%%%%%%%%%%%%%%%%%%%%%%%%%
\begin{wrapfigure}{r}{0.45\columnwidth}
%\begin{figure}
\centerline{\includegraphics[width=0.4 \columnwidth]{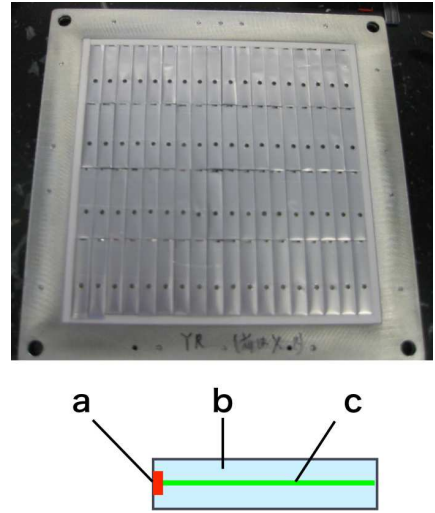}}
\caption{{\it top}:~A layer of the ScECAL prototype with 72 scintillator strips enveloped in reflector films. {\it bottom}:~A schematic
 of a scintillator strip unit with; a. PPD, b. scintillator, c. WLS fiber. }\label{fig:modulepict}
\end{wrapfigure}
%\end{figure}
%%%%%%%%%%%%%%%%%%%%%%%%%%%%%%%%%%%%%%%%
%%%%%%%%%%%%%%%%%%%%%%%%%%%%%%%%%%%%%%%%
Toward the particle flow approach, a granular electromagnetic calorimeter is mandatory \cite{MarkT}. 
To achieve 5 mm \kakeru\ 5 mm or  10 mm \kakeru\ 10 mm lateral granularity, the CALICE collaboration is developing a unique calorimeter concept using scintillator strips read out with Pixelated Photon Detectors (PPDs), 
where the scintillator strips in odd layers are orthogonal with respect to those in the even layers, having tungsten absorber layers in 
between the layers \cite{LOI}.
A prototype module of such a scintillator electromagnetic calorimeter (ScECAL) has been constructed and tested  in the period August-September 2008 and April-May 2009 at the FNAL meson beam test facility \cite{VCI}.
Figure~\ref{fig:modulepict} shows a picture of a layer and a schematic of the scintillator strip unit.
The scintillator strip that is hermetically enveloped with a reflector film has a width, length, and thickness of 10 mm, 45 mm, and 3 mm.
Therefore, an effective 10 mm \kakeru\ 10 mm segmentation is expected for this prototype.
The scintillation photons are collected by a wavelength shifting (WLS) fiber, inserted centrally, along the longitudinal direction of each strip and are read out with
a PPD provided by Hamamatsu Photonics KK, ``1600-pixel MPPC' \cite{HPK}.
The ScECAL prototype has a transverse area of 180 mm \kakeru\ 180 mm and 30 scintillator layers.
The thickness of 3.5 mm of 30 absorber layers lead to a total radiation length of 21.3 $X_0$.
%The 3.5 mm thick tungsten layers in between sensor layers lead to a total radiation length of 21.3 $X_0$. 

In the test beam experiment periods the ScECAL prototype was exposed at FNAL to electron and hadron beams up to 32 GeV/$c$, together with the analog 
scintillator hadron calorimeter \cite{HCAL} and the Tail Catcher \cite{TCMT} to evaluate their combined performance. % in the CALICE test beam activities.
The construction and basic performance of the ScECAL in those test beams is reported in \cite{VCI}.

Since the PPD is a semiconductor sensor, its sensitivity is influenced by temperature fluctuations.
In these proceedings we show how a temperature correction for such fluctuations can be achieved and how it performs well.

\section{Experiment and analysis}

\subsection{Beam and temperature environments }

\begin{figure}[!h]
\centerline{\includegraphics[width=0.7 \columnwidth]{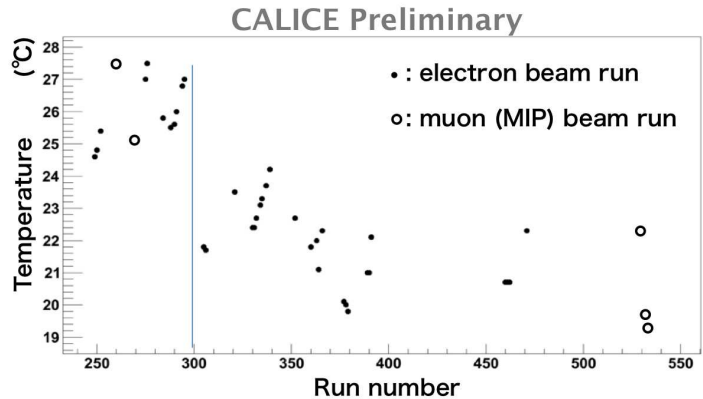}}
\caption{Temperature of the ScECAL surface for all electron and muon runs. 
The temperature is measured at two orthogonal positions of the ScECAL surface. 
A vertical blue line shows the run time where the temperature condition was drastically changed.}\label{fig:temperature}
%\end{wrapfigure}
\end{figure}

Temperature effects clearly appear on the spectra of electron energy scans. 
Although both 2008 and 2009 we took electron energy scan data, the temperature measurement system of the ScECAL malfunctioned in 2008.
In addition, as we had large temperature fluctuations, 19\,\celd\ - 27\,\celd, at our experimental hall during 2009 data taking, we use only 2009 data in this discussion.
The beam momenta are 2, 4, 8, 12, 15, 20, 30, 32 GeV/\SOL.
The measured temperature of the average of two positions on the ScECAL surface for each run is plotted in Figure~\ref{fig:temperature}. 
%%%%%%%%%%%%%%%%%%%%%%%%%%%%%%%%%%%%%%%%%%%%%%%%%
%%%%%%%%%%%%%%%%%%%%%%%%%%%%%%%%%%%%%%%%%%%%%%%%
%\begin{wrapfigure}{r}{1.4\columnwidth}

\subsection{Detector calibration with minimum ionizing particles}\label{sec:Calibration}
The key issue of the temperature correction of the sensitivity is the correction of the detector calibration with minimum ionizing particles (MIP).
The temperatures of MIP runs can be seen as open circles in Figure~\ref{fig:temperature}.
The muon-tuned beam contains almost no electrons or pions, because of the iron dump put in the beam line upstream of our experiment's site.
Therefore, the MIP events are only required to have the same lateral position of scintillator  "hit" of at least 10 on both even and odd layers.

\if0
\begin{wrapfigure}{r}{0.5\columnwidth}
%\begin{figure}
\centerline{\includegraphics[width=0.45 \columnwidth]{./figs/distMIP.eps}}
\caption{{\it top}:A layer of ScECAL prototype with 72 scintillator strips enveloped in reflector films. {\it bottom}:
Schematics of a scintillator strips with a. PPD, b. scintillator, c. WLS fiber. }\label{fig:modulepict}
\end{wrapfigure}
%\end{figure}
\fi

The MIP response for each channel is obtained by fitting a Landau function convoluted with a gaussian function to the distribution of ADC counts of the charge deposited by the MIP events.
The ADC/MIP conversion factor is the most probable value of the distribution.
%Figure~\ref{fig:MIPdist} shows a typical distribution of the deposited charge of the MIP events on a channel. 

\subsection{Temperature dependence of the ADC/MIP conversion factors}

%%%%%%%%%%%%%%%%%%%%%%%%%%%%%%%%%%%%%%%%%
%%%%%%%%%%%%%%%%%%%%%%%%%%%%%%%%%%%%%%%%%%%%%
\begin{wrapfigure}[21]{r}{0.48\columnwidth}
%\begin{figure} %%%%%%%%%% 0.38
\centerline{\includegraphics[width=0.43 \columnwidth]{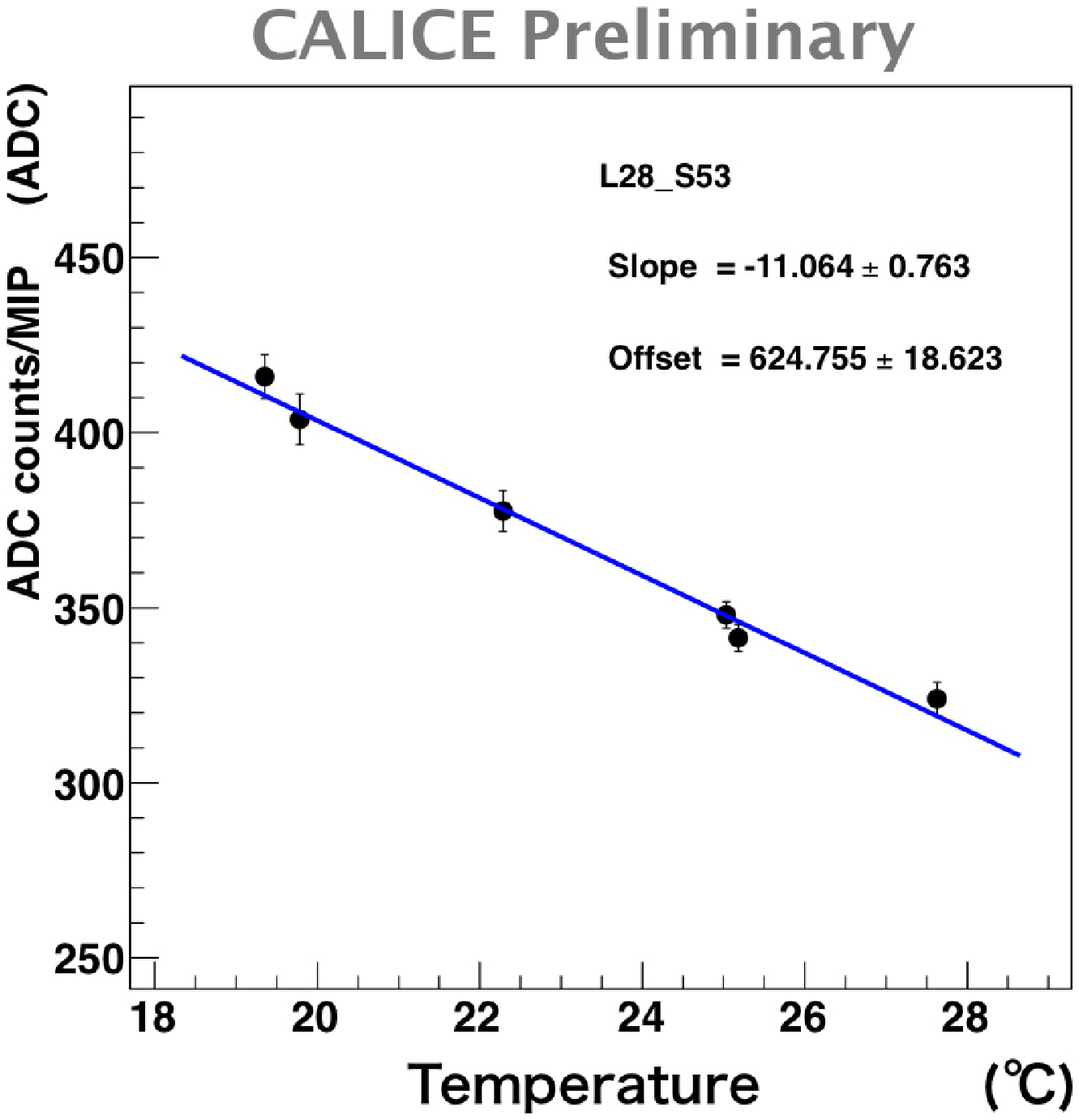}}
\caption{%typical 
Temperature dependence of $c_{ls}^{\mathrm{MIP}}$ and the result  of a linear fit.% to six temperature data. 
}\label{fig:mip_temp}
\end{wrapfigure}
%\end{figure}
%%%%%%%%%%%%%%%%%%%%%%%%%%%%%%%%%%%%%%%%%%%%%
%%%%%%%%%%%%%%%%%%%%%%%%%%%%%%%%%%%%%%%%%%%%%%

An ADC/MIP conversion factor estimated in Section \ref{sec:Calibration} should be a function of temperature during the data acquisition. 
The main temperature correction for a channel is achieved by using this temperature function of ADC/MIP conversion factor instead of a constant
 ADC/MIP conversion factor.
In 2009 a total of six MIP runs are taken at different temperatures ranging from 19\,\celd\ to 27\,\celd.
Figure~\ref{fig:mip_temp} shows a typical behavior of the ADC/MIP conversion factor for a channel as a linear function of the temperature. 
Temperature data were taken every minute and the temperature on the horizontal axis on Figure~\ref{fig:mip_temp} is the average of temperatures 
of every event in a run.
The function of ADC/MIP conversion factors obtained as a result of a linear fit is: %in Figure~\ref{fig:mip_temp} is:

\begin{equation}
c_{ls}^{\mathrm{MIP}}(T) = c_{ls}^{\mathrm{MIP}}(T_0) + \frac{dc_{ls}^{\mathrm{MIP}} }{dT},
\end{equation}

\noindent
%%%%%%%%%%%%%%%%%%%%%%%%%%%%%%%%%%%%%%%%%%%%%%%%%%
%%%%%%%%%%%%%%%%%%%%%%%%%%%  4  %%%%%%%%%%%%%%%%%%%
\begin{wrapfigure}[17]{r}{0.48\columnwidth}
%\begin{figure}
\centerline{\includegraphics[width=0.43 \columnwidth]{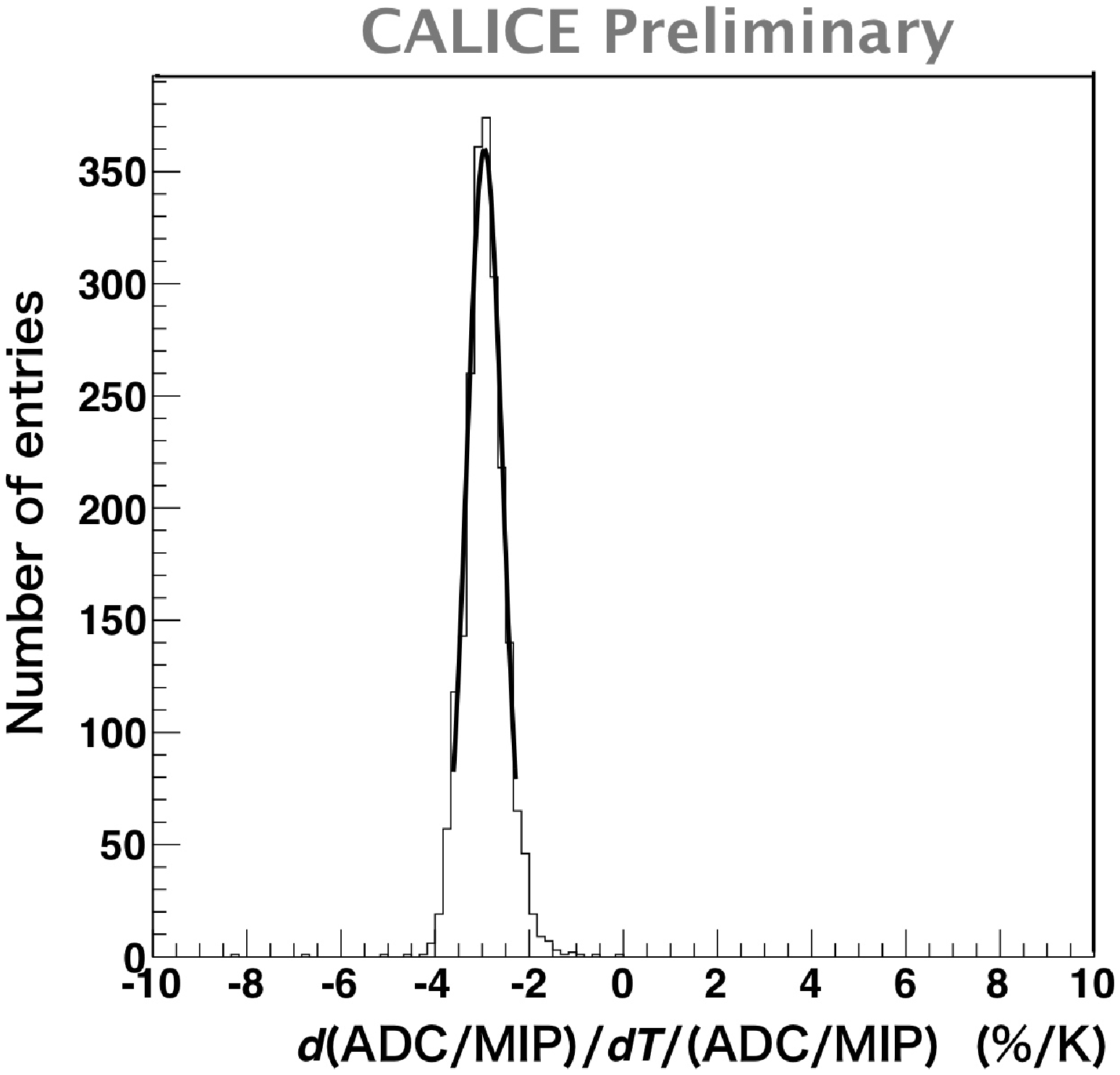}}
\caption{Distribution of $dc_{ls}^{\mathrm{MIP}}/dT/c_{ls}^{\mathrm{MIP}}$. 
The mean is -2.95$\pm$0.01\%/K and the $\sigma$ of from the fit is %standard deviation of fitted Gaussian is 
0.405\%/K . }\label{fig:mipslope}
\end{wrapfigure}
%%%%%%%%%%%%%%%%%%%%%%%%%%%%%%%%%%%%%%%%%%%%%%%%%%
%%%%%%%%%%%%%%%%%%%%%%%%%%%   4    %%%%%%%%%%%%%%%%%%%
%\end{figure}
where $c_{ls}^{\mathrm{MIP}}(T)$ is the ADC/MIP conversion factor for layer $l$ and strip number $s$ as a function of $T$, $dc_{ls}^{\mathrm{MIP}}/dT$ is the slope of the function, and $T_0$ is a fixed temperature for the offset value of this function. %, likely $T_0 = 0$\celd\ as usual. 
%Figure~\ref{ADCMIPdist}(a) shows the distribution of $c_{ls}^{\mathrm{MIP}}(T = 20$\celd$)$ and 
Figure~\ref{fig:mipslope} shows the distribution of 
$dc_{ls}^{\mathrm{MIP}}/dT/c_{ls}^{\mathrm{MIP}}$(20\celd). %  of over 2157 channels.
The temperature correction for each event for channel $l,s$ is done by using $c_{ls}^{\mathrm{MIP}}(T)$ with $T$ corresponding to the event data taking.

%Calibration by using $c_{ls}^{\mathrm{MIP}}(T)$ with $T$  corresponding to the event data taking makes temperature correction for each event for this
%channel $ls$. 

\if0
%\begin{wrapfigure}{r}{1.4\columnwidth}
\begin{figure}[!h]
\centerline{\includegraphics[width=0.6 \columnwidth]{./figs/mipslopeH.eps}}
\caption{Distribution of  $c_{ls}^{\mathrm{MIP}}(T = 20$\celd$)$. The mean value is 176 adccounts (a), and 
distribution of $dc_{ls}^{\mathrm{MIP}}/dT/c_{ls}^{\mathrm{MIP}}$. Mean value is -2.95$\pm$0.01\%/K and the standard deviation of result of Gaussian fit is 
0.405\%/K (b). }\label{fig:tm}
%\end{wrapfigure}
\end{figure}
\fi

\subsection{Energy spectra}
The energy spectrum for an incident beam momentum is the distribution of the sum of deposited charges of all channels in an event, which converted into
the number of corresponding MIPs estimated by: % using the ADC/MIP conversion factor $c_{l's'}^{\mathrm{MIP}}(T)$, where $T$ is temperature at the data taking of this hit.
%The sum of deposited charge $E_{\mathrm{total}}$ is:
\begin{equation}
E_{\mathrm{total}} = \sum_{l=1}^{30} \sum_{s=1}^{72}\frac{E'_{ls}}{c_{ls}^{\mathrm{MIP}}(T)},
\end{equation}
where $E_{\mathrm{total}}$ is the sum of deposited charges converted into the number of corresponding MIPs, and $T$ is the temperature corresponding to
 this event, and $E'_{ls}$ is the deposited charge in channel $l,s$ applied the PPD saturation correction.
The PPD saturation correction is achieved  as:
\begin{equation}
E'_{ls} = - n_{\mathrm{pix}} R_{\mathrm{ADC-photon},ls}\log\Big(1-\frac{N_{\mathrm{ADC},ls}/R_{\mathrm{ADC-photon},ls}}{n_{\mathrm{pix}}}\Big),
\end{equation}
where
$n_{\mathrm{pix}}$ is the effective number of pixels discussed in \cite{VCI}, 
$N_{\mathrm{ADC},ls}$ is the number of ADC counts as the response of strip $l,s$, and 
$R_{\mathrm{ADC-photon},ls}$ is the ratio of ADC counts to the number of detected photons for the channel $l,s$ obtained by using the method
descrived in  \cite{VCI}.  

To sum up the deposited charges% with respect to events
, the same criteria of event selection as in \cite{VCI} is implemented.

\section{Result}
\subsection{Linearity of the energy response}

\begin{figure}[!h]
\centerline{\includegraphics[width=1.0  \columnwidth]{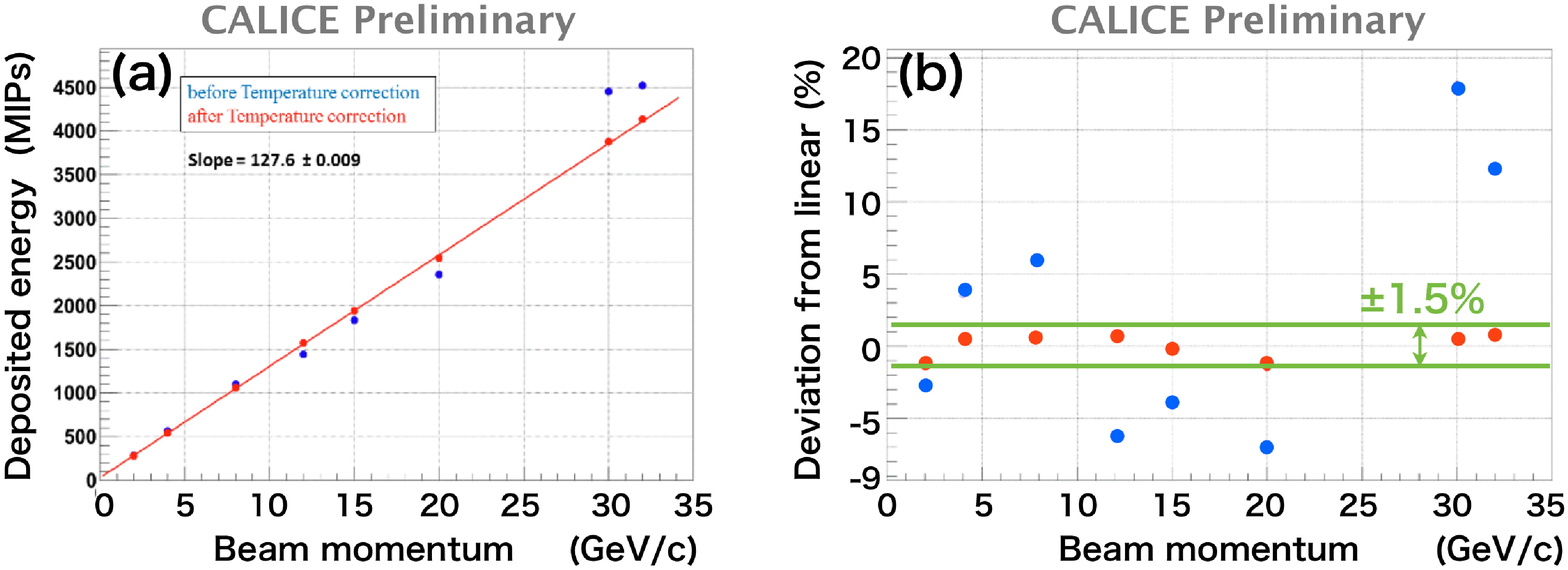}}
\caption{(a) ScECAL response to electron beams. Blue and red dots show the results without and with the temperature correction, respectively. The solid line shows the result of a linear fit to the red dots. (b) Deviation from the result of the linear fit shown in (a). }\label{fig:linearity}
%\end{wrapfigure}
\end{figure}

Figure~\ref{fig:linearity}(a) shows the deposited energy in the ScECAL prototype as a function of the beam momentum.
Blue dots show the results without temperature correction and red dots show the results with temperature correction.
The significant improvement by the temperature correction becomes clear when we see the deviation from a linear fit in Figure~\ref{fig:linearity}(b).
The deviation from the linear fit is improved from greater than $\pm$10\% to less than $\pm$1.5\%.

%%%%%%%%%%%%%%%%%%%%%%%%%%%%%%%%%%%%%%%%%%%%%
%%%%%%%%%%%%%%%%%%%%%%%%%%%%%%%%%%%%%%%%%%%%%%
%\begin{wrapfigure}{r}{1.4\columnwidth}
%%%%%%%%%%%%%%%%%%%%%%%%%%%%%%%%%%%%%%%%%%%%%
%%%%%%%%%%%%%%%%%%%%%%%%%%%%%%%%%%%%%%%%%%%%%%

\subsection{Resolution of the energy response}
\begin{wrapfigure}[20]{r}{0.6\columnwidth}
%\begin{figure}
\centerline{\includegraphics[width=0.5 \columnwidth]{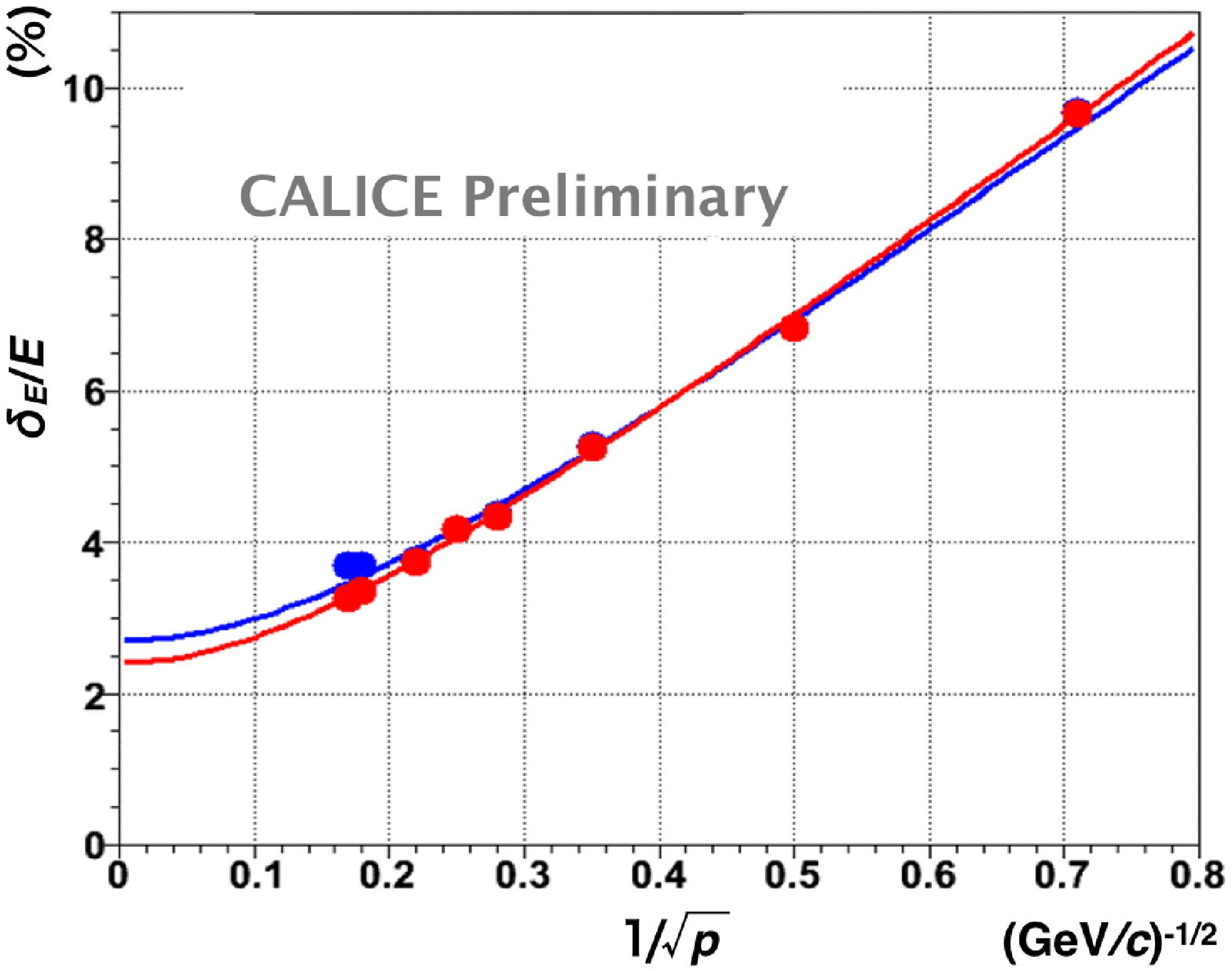}}
\caption{Energy resolutions of the ScECAL for electron beams. Blue and red dots show the result without and with temperature correction, respectively. 
Blue and red curves show the result of  fits of a quadratic parametrization to the resolution with and without temperature correction, respectively.}\label{fig:resolution}
\end{wrapfigure}

Figure~\ref{fig:resolution} shows the energy resolution as a function of $1/\sqrt{P_{\mathrm{beam}}}$, where $p$ is the beam momentum. 
Blue dots show the result  without temperature correction and red dots show the result with temperature correction.
The curves are the results of fits with a quadratic parametrization of the resolution:

\begin{equation}
\frac{\sigma_E}{E} = \sigma_{\mathrm{constant}} \oplus \sigma_{\mathrm{stochastic}}\frac{1}{\sqrt{E_{\mathrm{beam}}}},
\end{equation}

%%%%%%%%%%%%%%%%%%%%%%%%%%%%%%%%%%%%%%%%%%%%%
%%%%%%%%%%%%%%%%%%%%%%%%%%%%%%%%%%%%%%%%%%%%%%
%\end{figure}
%%%%%%%%%%%%%%%%%%%%%%%%%%%%%%%%%%%%%%%%%%%%%
%%%%%%%%%%%%%%%%%%%%%%%%%%%%%%%%%%%%%%%%%%%%%%
\noindent where $E$ is 
the deposited energy in the ScECAL, $\sigma_E$ the energy resolution, $E_{\mathrm{beam}}$ the beam energy, and 
$\sigma_{\mathrm{constant}}$ and $\sigma_{\mathrm{stochastic}}$ the constant and the stochastic term of the energy resolution parametrization, 
respectively.
After temperature correction, 
the stochastic and the constant term of the energy resolution parametrization are $13.16\pm0.05\%$ and $2.32 \pm 0.02\%$, respectively, 
where the uncertainties are only statistical uncertainties.

\section{Discussion}

Fluctuations of the measured energy due to the temperature fluctuations do not only come from the temperature dependence of 
the PPD but also come from other effects.
For example, the scintillation process is also affected by temperature fluctuation.
The temperature correction implemented in this study totally includes effects of those fluctuations.
The significant improvement  of the linearity of the ScECAL response by using this temperature correction shows that the 
main effects are taken into account. %
The remaining temperature correction we need is the fluctuation that comes from the PPD saturation correction by using the number of fired pixels.
%However, temperature dependence of PPD saturation correction by using the number of fired pixel cannot be corrected with this method.  
Although we can expect that the fluctuation coming from the saturation correction of the PPD affected by temperature changes is small, 
we will also implement a temperature correction for this effect by using the temperature dependence of the PPD gain in the near future.

There are two positions on the surface of the detector for the temperature measurements.
The temperature sensors are positioned orthogonally to each other.
Since the difference of the measured temperatures of those two positions is always less than
%Measurement positions of temperature are two on the surface of the detector, and each temperature sensor is positioned orthogonally from the other.
%From the fact that difference of temperature of those two position position is always less than
 0.3\, \celd, 
 we assume that the surrounding atmosphere 
has a uniform temperature by using a fan near the detector.
Although we do not have data of the temperature distribution inside of the detector so far, we will investigate it with the electron currents of the 
PPDs and the fluctuation rate of the surrounding temperature in the next prototype.

\section{Summary}

The temperature correction of the MIP calibration for the energy measurement was implemented in the analysis of data of the ScECAL prototype.
A significant improvement of the linearity of the calorimeter response for 2 - 32 GeV electron was confirmed.
The maximum deviation from the linear fitting result is $\pm$ 1.5\% with this temperature correction, while it is larger than $\pm$10\% without
temperature correction.

\section{Acknowledgments}
The author gratefully thanks the CALICE members who contributed  to the FNAL test beam experiment, especially for the analog hadron calorimeter group 
for the construction of the data taking system also for this ScECAL.
The author also would like to express special thanks to the Fermilab accelerator staff for providing a good quality beam. 
This study is supported in part by the Creative Scientific Research Grants No.~18GS0202 of the Japan Society for Promotion of Science (JSPS)
and the World Class University project (WCU) by Korea National Research Foundation.

% ****************************************************************************
% BIBLIOGRAPHY AREA
% ****************************************************************************

\begin{footnotesize}
% IF YOU DO NOT USE BIBTEX, USE THE FOLLOWING SAMPLE SCHEME FOR THE REFERENCES
% ----------------------------------------------------------------------------

% ----------------------------------------------------------------------------

\end{footnotesize}

% ****************************************************************************
% END OF BIBLIOGRAPHY AREA
% ****************************************************************************

\end{document}